\documentclass[aps,prd,
showpacs,showkeys,amsmath,amssymb,
oneside,onecolumn,notitlepage,floatfix,superscriptaddress]{revtex4-1}
\usepackage{graphicx}
\usepackage{times}
\usepackage{verbatim}
\usepackage{color}
\usepackage{cancel}
\usepackage[normalem]{ulem}
\usepackage{multirow}
\usepackage{braket}
\usepackage{bm}
\usepackage{hyperref}
\usepackage{tabularx}
\usepackage[capitalise]{cleveref}
\usepackage[subpreambles=true,sort]{standalone}
\usepackage{slashed}

\newcommand{\Dubna}{\affiliation{Joint~Institute~for~Nuclear~Research, Dubna, Moscow~Region, Russia}}

\begin{document}
\title{On coherent neutrino and antineutrino scattering off nuclei}
\author{Vadim~A.~Bednyakov}\Dubna
\author{Dmitry~V.~Naumov}\Dubna
\date{\today}

\begin{abstract}
Neutrino-nucleus $\nu A\to \nu A$ and antineutrino-nucleus $\bar\nu A\to \bar\nu A$ interactions, when the nucleus conserves its integrity, are discussed with coherent (elastic) and incoherent (inelastic) scattering regimes taken into account.
In the first regime the nucleus remains in the same quantum state after the scattering and the cross-section depends on the quadratic number of nucleons. 
In the second regime the nucleus changes its quantum state and the cross-section has an essentially linear dependence on the number of nucleons. 
The coherent and incoherent cross-sections are driven by a nuclear nucleon 
form-factor squared $|F|^2$ term and a $(1-|F|^2)$ term, respectively. 
One has a smooth transition between the regimes of coherent and incoherent (anti)neutrino-nucleus scattering. 
Due to the neutral current nature these elastic and inelastic  processes are indistinguishable if the nucleus recoil energy is only observed. 
One way to separate the 
coherent signal from the incoherent one is to register $\gamma$ quanta
from deexcitation of the nucleus excited during the incoherent scattering. 
Another way is to use a very low-energy threshold detector and collect data at very low recoil energies, where the incoherent scattering is vanishingly small.
In particular, for ${}^{133}\text{Cs}$ and neutrino energies of 30--50 MeV the incoherent cross-section is about 15-20\% of the coherent one. 
Therefore, the COHERENT experiment (with ${}^{133}\text{Cs}$) has measured  
the coherent elastic neutrino nucleus scattering (CE$\nu$NS)
with the inelastic admixture at a level of 15-20\%, if the excitation $\gamma$ quantum escapes its detection.  
\end{abstract}

\maketitle 

After Freedman's paper~\cite{Freedman:1973yd} it was confirmed
~\cite{Drukier:1983gj,Barranco:2005yy,Patton:2012jr,Papoulias:2015vxa} that in the Standard Model the cross-section of elastic neutrino scattering off a nucleus
is enhanced with respect to neutrino scattering off a single nucleon.
The amplification factor for  a spinless even-even nucleus is
$\left|g_V^nN F_n(\bm{q})+g_V^pZF_p(\bm{q})\right|^2\simeq N^2 (g_V^n)^2|F_n(\bm{q})|^2$,
giving the {coherent} $\nu A$-scattering cross-section in the well-known form
\cite{Freedman:1973yd,Drukier:1983gj,Barranco:2005yy,Patton:2012jr,Papoulias:2015vxa,Smith:1985mta,Jachowicz:2001jr,Divari:2010zz,McLaughlin:2015xfa,Vergados:2009ei,Papavassiliou:2005cs,Divari:2012zz}
\begin{equation}\label{eq:CS-enhancement_factor_naive}
\frac{d\sigma_\text{coh}}{dT_A}   \approx \frac{G_F^2 m_A}{\pi}\left(1-\frac{T_A}{T_A^\text{max}}\right)|F_n|^2 \left(g_V^n\right)^2N^2. 
\end{equation} 
Here $T_A$ is the kinetic energy of the scattered nucleus, $m_A$ is the nucleus mass, 
$\bm{q}$ is the momentum transfer, $G_\text{F}$ is the Fermi constant, 
$Z$ and $N$ are the numbers of protons and neutrons, $g_V^{p/n}$ are the proton/neutron couplings of the nucleon vector current, and $F_{p/n}(\bm{q})$ are the proton/neutron form-factors of the nucleus.
The form-factors vanish as $|\bm{q}|\to\infty$ and approach unity ($F_{p/n}(\bm{q})=1$) if 
$|\bm{q}|R_A\ll 1$, where $R_A$ is the radius of the nucleus. 
The coherency requirement reads as $|\bm{q}|R\ll 1$.

Freedman used the termin ''coherent neutrino-nucleus scattering'' (CNNS) 
~\cite{Freedman:1973yd} to emphasize the fact that the dependence of the corresponding cross-section is quadratic in terms of the number of nucleons. 

The importance of the CNNS was demonstrated for a number of observables in astrophysics, like stellar collapse~\cite{Wilson:1974zz,Freedman:1977xn}, and Supernovae~\cite{Bernabeu:1975tw,Rombouts:1997vm,Divari:2012zz,Divari:2012cj}, 
in studies of physics beyond the Standard Model (SM) ~\cite{Papavassiliou:2005cs,Barranco:2005yy,Scholberg:2005qs,deNiverville:2015mwa,Esteban:2018ppq,Abdullah:2018ykz,Farzan:2018gtr,Billard:2018jnl,Denton:2018xmq,Ge:2017mcq,Kosmas:2017tsq,Canas:2018rng,AristizabalSierra:2018eqm}, 
and in investigation of the nuclear structure 
~\cite{Engel:1991wq,Amanik:2009zz,Amanik:2007ce,Patton:2013nwa,Patton:2012jr,Cadeddu:2017etk}.
Due to the neutral-current nature an observation of $\nu$-oscillations with CNNS could be evidence for sterile neutrino(s) \cite{Formaggio:2011jt,Anderson:2012pn}. 
Coherent scattering off atomic systems was studied in~\cite{Gaponov:1977gr,Sehgal:1986gn}. 
There are some experimental proposals to observe the CNNS
~\cite{Lewis:1979mu,Drukier:1983gj,Horowitz:2003cz,Giomataris:2005fx,Wong:2005vg,Vergados:2009ei,Sangiorgio:2012vsa,Brice:2013fwa,Kopylov:2013zda,Kopylov:2014xra,Agnolet:2016zir,Aguilar-Arevalo:2016khx,Moroni:2014wia,Belov:2015ufh,Tayloe:2017edz,Billard:2016giu}.
This process is an unavoidable background in sensitive direct dark matter searches ~\cite{Wong:2010zzc,Anderson:2011bi,Gutlein:2014gma,Bednyakov:2015uoa,Fallows:2018ika}.
Due to the CNNS one expects to reduce {significantly the size} of a neutrino detector. It  would help to develop neutrino-based applied research (non intrusive monitoring of nuclear reactors, etc).

The difficulty in observing CNNS lies in the detection of scattered nuclei with low kinetic energy of the order of a few keV. 
This nuclear recoil energy is the only measurable CNNS signature.  
Detection of neutrinos (with $E_\nu<$50~MeV) via CNNS is a challenge. 

The first experimental evidence for CNNS  was reported in 2017 by the COHERENT Collaboration~
\cite{Bolozdynya:2012xv,Akimov:2015nza,Collar:2014lya}, who used the CsI[Na] scintillator exposed to  neutrinos with energies of 
tens of MeV produced by the Spallation Neutron Source (SNS) at the Oak Ridge National Laboratory~
\cite{Akimov:2017ade,Akimov:2018vzs,Akimov:2018ghi}.
The COHERENT energy threshold was {\em 5} keV (for caesium).
At these energies the momentum transfer $\bm{q}$ is large enough to break the  condition $|\bm{q}|R_A \ll 1$.
For example, energy deposits observed in~\cite{Akimov:2017ade} correspond to $1<|\bm{q}|R_A<2.7$, 
and the pure elastic cross-section should be suppressed.
At higher energies the elastic cross-section (given in~\cref{eq:CS-enhancement_factor_naive})
vanishes (due to form-factors), but the neutrino-nucleus interaction probability, obviously, does not vanish and 
must be determined by some inelastic interaction (absent in~\cref{eq:CS-enhancement_factor_naive}). 
In general, the corresponding cross-section should be given by a sum of elastic and inelastic cross-sections, similar to the theory of the scattering of $X$ rays~\cite{Waller119} and electrons~\cite{Morse1932} off an atom and of slow neutrons off matter constituents~\cite{PhysRev.95.249}.

In our previous paper 
\cite{Bednyakov:2018mjd} a theoretical framework allowing for elastic and inelastic neutrino-nucleus  scattering in the process 
$\nu A \to \nu A^{(*)}$ was developed on the basis of calculations from first principles. 
The possibility that the  internal quantum state of a nucleus can be modified after an interaction is labeled by the $(*)$ superscript.

\smallskip
In this paper new results for the neutrino-nucleus, $\nu A \to \nu A^{(*)}$, 
and antineutrino-nucleus, $\bar\nu A \to \bar\nu A^{(*)}$, 
elastic and inelastic scattering processes obtained within the theoretical framework
of \cite{Bednyakov:2018mjd} are presented and briefly discussed.

Neutrinos and antineutrinos with energies below tens of MeV predominately conserve the integrity of nucleons in neutrino-quark interactions with $Z^0$-boson exchange, 
allowing usage of an effective (anti)neutrino-nucleon interaction in the form 
$\displaystyle \mathcal{L}(x) = \frac{G_F}{\sqrt{2}} L_\mu(x)H^\mu(x).$
Here  
$\displaystyle L_\mu(x)= \overline{\psi}_\nu(x)\gamma_\mu(1-\gamma_5)\psi_\nu(x)$ and 
$\displaystyle H^\mu(x) =\sum_{f=n,p}\overline{\psi}_f(x) O^\mu_f\psi_f(x)$
are the weak currents of (anti)neutrinos and nucleons, respectively, 
$O^\mu_f =\gamma^\mu\left(g_V^f-g_A^f\gamma_5\right)=\gamma^\mu\left(g_L^f(1-\gamma_5)+g_R^f(1+\gamma_5)\right)$, 
and left- and right-chirality couplings 
$g_{L/R}^{f} =\frac{1}{2}\left(g_V^{f}\pm g_A^{f}\right)$ are expressed in terms of the vector $g_V^{p/n}$ and axial $g_A^{p/n}$ couplings
with $g^{}_A=1.27\pm 0.003$
\cite{Tanabashi:2018oca}
\begin{equation}
g_V^p = \frac{1}{2}-2\sin^2\theta_W, \quad  g_V^n =-\frac{1}{2}, \quad 
g_A^p =  \frac{g^{}_A}{2}, \quad  g_A^n =-\frac{g^{}_A}{2}.
\label{eq:sm_couplings}
\end{equation}
In \cite{Bednyakov:2018mjd} the SM coupling values were used (with $g^{}_A\equiv 1$).

As demonstrated in \cite{Bednyakov:2018mjd},  
the neutrino-nucleus scattering $\nu(k)+A\!\to\!\nu(k')+ A^{(*)}$ cross-section  is a sum of incoherent and coherent terms. 
Following \cite{Bednyakov:2018mjd}, one has this sum for the antineutrino-nucleus scattering, 
$\bar\nu(k)+A\!\to\!\bar\nu(k')+A^{(*)}$, as well
\begin{equation}
\label{eq:main-cross-section}
\frac{d\sigma^{\nu/\bar\nu}}{dT_A} = \frac{d\sigma^{\nu/\bar\nu}_\text{incoh}}{dT_A} + \frac{d\sigma^{\nu/\bar\nu}_\text{coh}}{dT_A}.
\end{equation}
Here and below the left and upper symbols stand for neutrinos, and the right and lower symbols stand for antineutrinos.

The incoherent (anti)neutrino-nucleus scattering cross-section in (\ref{eq:main-cross-section}) is
\begin{eqnarray}\nonumber\label{eq:sigma-inc1}
\frac{d{\sigma}^{\nu/\bar\nu}_\text{incoh}}{dT_A}
&=&
\frac{4G_F^2 m_A}{\pi}\sum_{f=p,n}g^f_\text{inc}\left[1-|F_f(q^2)|^2\right] \left[
A^f_{\mp}(g^{f}_{R/L})^2\,\,\frac{s(1-y)^2-m^2(1-y)}{s-m^2}+
\right.\\&& \left.\nonumber\qquad\qquad\qquad\qquad
+A^f_{\pm}\left\{\left(g^{f}_{L/R}-g^{f}_{R/L}\frac{ym^2}{s-m^2}
\right)^2 +(g^f_{R/L})^2 \,\, \frac{ym^2[s(1-y)-m^2]}{(s-m^2)^2}\right\} \right]=
\\[-8pt]\\\nonumber   
&=&\frac{2G_F^2 m_A}{\pi}\sum_{f=p,n} g^f_\text{inc}\left[1-|F_f(q^2)|^2\right]
\left[A^f\left\{(g^f_{L/R})^2+(g^f_{R/L})^2(1-y)^2- 2 g^f_Lg^f_R\, \frac{ym^2}{s-m^2}\right\}+
\right. \\&& \left.  \nonumber \qquad\qquad\qquad\qquad
+(\pm{\Delta A^f})\left\{g^f_{L/R}-g^f_{R/L}(1-y)\right\}\left\{g^f_{L/R}+g^f_{L/R}\left(1-y\frac{s+m^2}{s-m^2}\right)\right\}\right].
\end{eqnarray}

The coherent (anti)neutrino-nucleus scattering cross-section in (\ref{eq:main-cross-section}) is 
\begin{eqnarray}\label{eq:sigma-coh1}\nonumber
\frac{d{\sigma}^{\nu/\bar\nu}_\text{coh}}{dT_A}&=&\frac{4G_F^2 m_A}{\pi}
\left(1-\frac{T_A}{T_A^{\max}}\right) \left| \sum_{f=p,n} \sqrt{g^f_\text{coh}} F_f(q^2)
\left[A^f_{\pm}\left\{g^f_{L/R}+ g^f_{R/L}\,\frac{my}{\sqrt{s}+m}\right\}
+ A^f_{\mp}\,g^f_{R/L}\left\{1-\frac{\sqrt{s}y}{\sqrt{s}+m}\right\}\right]\right|^2=
\\
&=&
\frac{G_F^2 m_A}{\pi}\left(1-\frac{T_A}{T_A^{\max}}\right)
\left|G^{\nu/\bar\nu}_V(q^2)+G^{\nu/\bar\nu}_A(q^2)\right|^2, \qquad\text{where}
\\&&\qquad\qquad\qquad \nonumber
G^{\nu/\bar\nu}_V(q^2)=\sum_{f=p,n}\sqrt{g^f_\text{coh}}F_f(q^2) g^f_{V}\left[{A^f}\left(1-\frac{y\tau}{2}\right)+\frac{y}{2}(\pm{\Delta A^f})\right],
\\&&\qquad\qquad\qquad \nonumber
G^{\nu/\bar\nu}_A(q^2)=\sum_{f=p,n}\sqrt{g^f_\text{coh}}F_f(q^2)(\pm g^f_A)
\left[{A^f}\frac{y\tau}{2}+\left(1-\frac{y}{2}\right)(\pm{\Delta A^f})\right].
\end{eqnarray} 
Kinematic variables are $q^2=(k-k')^2$, $\displaystyle y=\frac{(p,q)}{(p,k)} \simeq \frac{T_A}{E_\nu}$, 
where $T_A$ is the kinetic energy of the nucleus.
One has $\displaystyle E_\nu = \frac{s-m^2}{2\sqrt{s}}$,
$\displaystyle \tau=\frac{\sqrt{s}-{m}}{\sqrt{s}+{m}}$, and $m$ is the nucleon mass. 
The total energy squared $s=(p+k)^2$ of the (anti)neutrino and the target nucleon is calculated assuming an effective momentum of the nucleon
\cite{Bednyakov:2018mjd}. 
In~Eqs.~(\ref{eq:sigma-inc1}) and (\ref{eq:sigma-coh1}) 
$A^f_\pm=(A^f\pm \Delta A^f)/2$, and $\displaystyle A^f=A^f_+ + A^f_-$, $\displaystyle \Delta A^f = A^f_+ - A^f_-$, or directly
$A^p=Z$, $A^n=N$, $\Delta A^p\equiv \Delta Z=Z_+-Z_-$, $\Delta A^n\equiv \Delta N=N_+-N_-$, where $Z_\pm$ and $N_\pm$ stand for the numbers of the protons and neutrons with the spin projection on the incident neutrino momentum axis equal to $\pm 1/2$.
Correction functions $g_{\text{inc}}^{p/n}$ and $g_{\text{coh}}^{p/n}$ are of the order of unity (for details, see \cite{Bednyakov:2018mjd}). 
For simplicity, in what follows these correction functions are omitted (or taken to be equal to one).  

If the target nucleus is unpolarized, the terms  proportional to $\Delta A^f$ in
(\ref{eq:sigma-inc1}) vanish after averaging, and for an unpolarized target 
the incoherent (anti)neutrino-nucleus scattering  cross-section is
\begin{equation}\label{eq:sigma-inc2}
\frac{d{\sigma}^{\nu/\bar\nu}_\text{incoh}}{dT_A}=\frac{2G_F^2 m_A}{\pi}
\sum_{f=n,p}\left[1-|F_f(q^2)|^2\right]A^f 
\left\{(g_{L/R}^f)^2+(g_{R/L}^f)^2(1-y)^2- 2 g_L^fg_R^f\frac{y m^2}{s-m^2}\right\}.
\end{equation}

Spin averaging in~(\ref{eq:sigma-coh1}) removes terms linear in $\Delta A_f$.
The formula of the spin-averaged coherent (anti)neutrino-nucleus scattering cross-section is
\begin{eqnarray}\label{eq:sigma-coh2}
\frac{d{\sigma}^{\nu/\bar\nu}_\text{coh}}{dT_A}
&=& \frac{G_F^2 m_A}{\pi}\left(1-\frac{T_A}{T_A^\text{max}}\right)\sum_{f,f'}F_fF^*_{f'}
\Bigg\{g_V^fg_V^{f'}\Bigg[A_f A_{f'}\left(1-\frac{y\tau}{2}\right)^2 + \Delta A_f \Delta A_{f'}\left(\frac{y}{2}\right)^2\Bigg]+
\\
&+&g_A^fg_A^{f'}\Bigg[A_f  A_{f'}\left(\frac{y\tau}{2}\right)^2+
\Delta A_f \Delta A_{f'} \left(1-\frac{y}{2}\right)^2 \Bigg]
+2g_V^f(\pm g_A^{f'}) \Bigg(A_fA_{f'}\left(1-\frac{y\tau}{2}\right)\frac{y\tau}{2}+\Delta A_f \Delta A_{f'}\frac{y}{2}\left(1-\frac{y}{2}\right)\Bigg)\Bigg\}.
\nonumber
\end{eqnarray}
Equations (\ref{eq:sigma-inc2}) and (\ref{eq:sigma-coh2}) can be further simplified 
if terms proportional to $y\approx 3\%E_\nu/(30\text{~MeV})$ 
and to $\Delta A_f \Delta A_{f'}$ are neglected. 
This can  be done either for a spinless nucleus or approximately for heavy nuclei with $\Delta A\ll A$.
Therefore, one has both coherent and incoherent terms in a rather simple form, 
which is the same for incoming neutrinos and antineutrinos
\begin{eqnarray}\nonumber\label{eq:for-both-Nu-and-ANu}
\frac{d\sigma_\text{coh}}{dT_A} &=& \frac{G_F^2 m_A}{\pi} \left(1-\frac{T_A}{T_A^\text{max}}\right)
 \sum_{f=n,p} {A^2_f} {| F_f |^2} (g_V^f)^2 ,
 \\[-10pt] \\ \nonumber 
\frac{d\sigma_\text{incoh}}{dT_A}&=&\frac{2G_F^2 m_A}{\pi} \sum_{f=n,p}{A_f} 
{\left(1-|F_f|^2\right)} \left((g_L^f)^2+(g_R^f)^2\right).
\end{eqnarray}
One can conclude from Eq. (\ref{eq:for-both-Nu-and-ANu}) that 
at the accepted accuracy level  
$\nu A\to \nu A^{(*)}$ and $\bar\nu A\to \bar\nu A^{(*)}$ interactions are identical.  
Finally, if terms proportional to $g_V^p$ are abandoned (due to $g_V^p\ll 1$),  
Eq.~(\ref{eq:sigma-coh2}) arrives at a well-known result given in 
Eq.~(\ref{eq:CS-enhancement_factor_naive}). 

A smooth transition between the coherent and incoherent regimes is the key feature of Eqs.~(\ref{eq:main-cross-section}), (\ref{eq:sigma-inc1}), and (\ref{eq:sigma-coh1}).
The elastic (coherent) interactions keeping the nucleus in the same quantum state lead to {quadratic enhancement} ($\propto A^2_f$) of the cross-section in terms of the number of nucleons 
and is simultaneously proportional to ${| F_f(\bm{q})|^2}$. 
The cross-section of the inelastic (incoherent) processes in which the quantum state of the nucleus is changed has the linear dependence ($\propto A_f$) on the number of nucleons
and is  simultaneously proportional to ${\left(1-|F_f(\bm{q})|^2\right)}$.  
Both terms in Eqs.~(\ref{eq:sigma-inc1}), and (\ref{eq:sigma-coh1}) 
are governed by the same $F_{f}(\bm{q})$. 
In the limit $\bm{q}\to 0$, $F_{f}(\bm{q}) \to  1$, and the contribution of the incoherent cross-section  
(see Eqs.~(\ref{eq:sigma-inc1}))
vanishes, while the coherent term totally dominates.
In the opposite limit of large $\bm{q}$, $F_{p/n}(\bm{q}) \to 0$, and the coherent cross-section vanishes (see Eqs.~(\ref{eq:sigma-coh1})), while the incoherent term dominates.
In general, both the coherent and incoherent scattering processes contribute.
In what follows, the results obtained with  
the Helm form-factors~\cite{PhysRev.104.1466} are presented. 

It is convenient to refer to the  cross-section integrated over the kinetic energy of the recoil nucleus
\begin{equation}
\label{eq:integral_xs}
\sigma(E_{\nu/\bar\nu})  = \int_{T_A^\text{min}} ^{T_A^\text{max}}\frac{d\sigma^{\nu/\bar\nu}}{dT_A}dT_A.
\end{equation}
This integral depends on the energy threshold $T_A^\text{min}$, unique for each detector.
\begin{figure}[h!]
\includegraphics[width=0.7\linewidth]{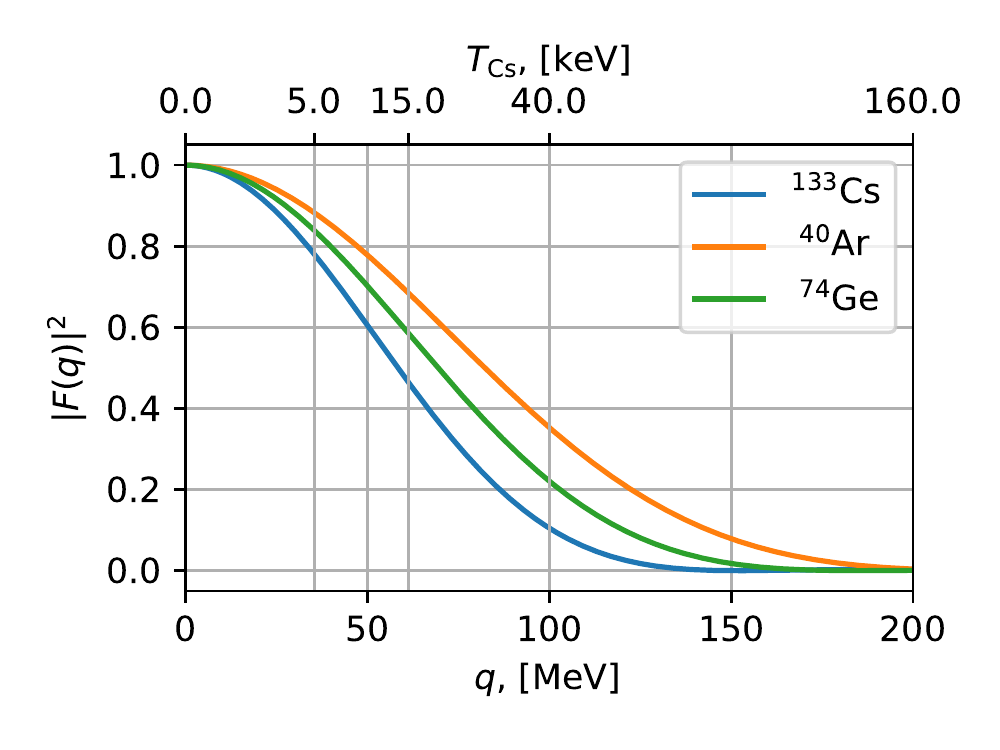}\vspace*{-20pt}
\caption{The Helm form-factor $F^\text{Helm}$~\cite{PhysRev.104.1466} as  a function of 
the three-momentum transfer $|\bm{q}|$ (bottom horizontal axis).
The upper horizontal axis corresponds to the kinetic energy of the ${}^{133}\text{Cs}$ nucleus.}
\label{fig:form-factors}
\end{figure}	

\begin{figure}[p]
\includegraphics[width=0.49\textwidth]{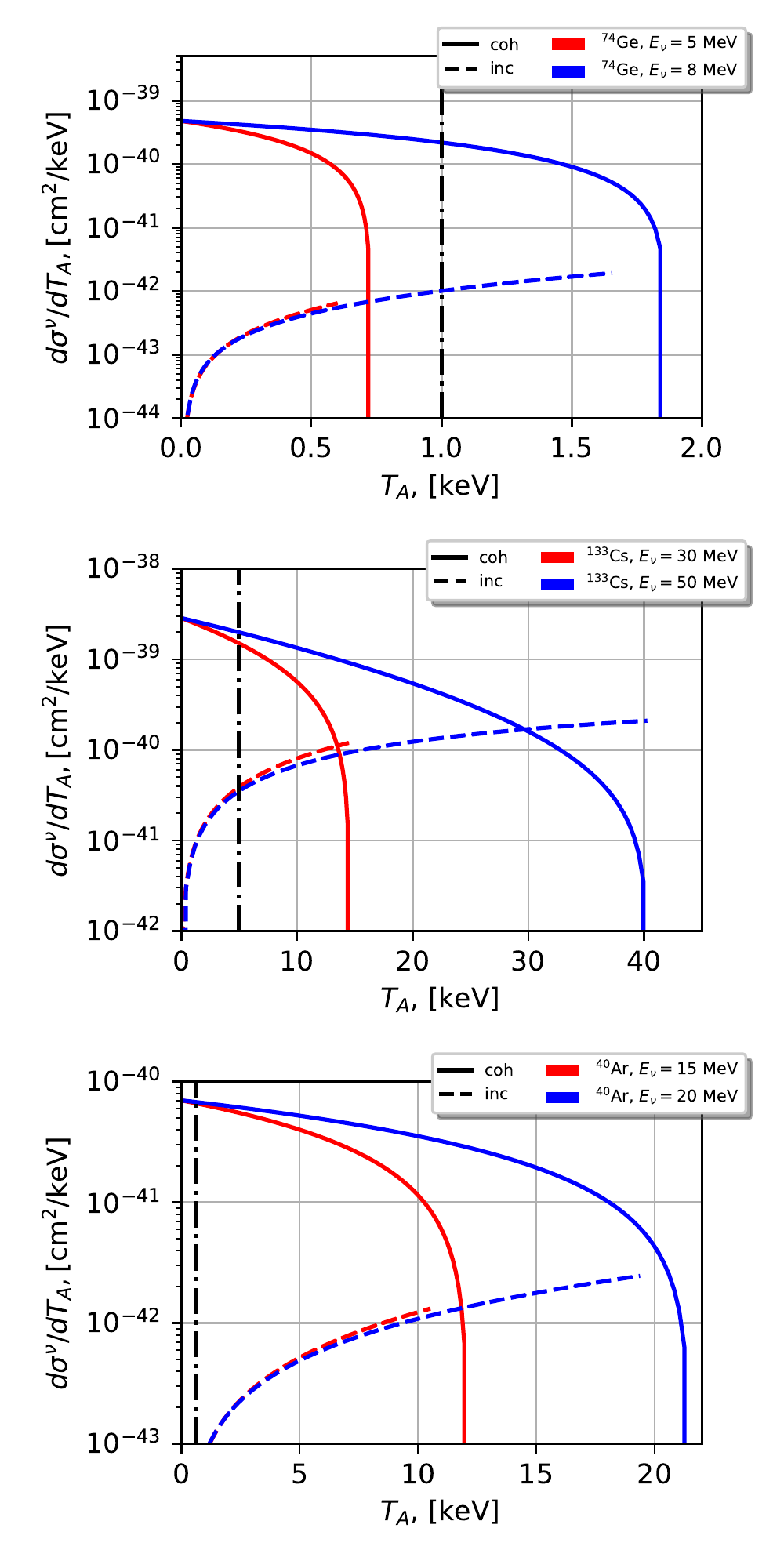}\
\includegraphics[width=0.49\textwidth]{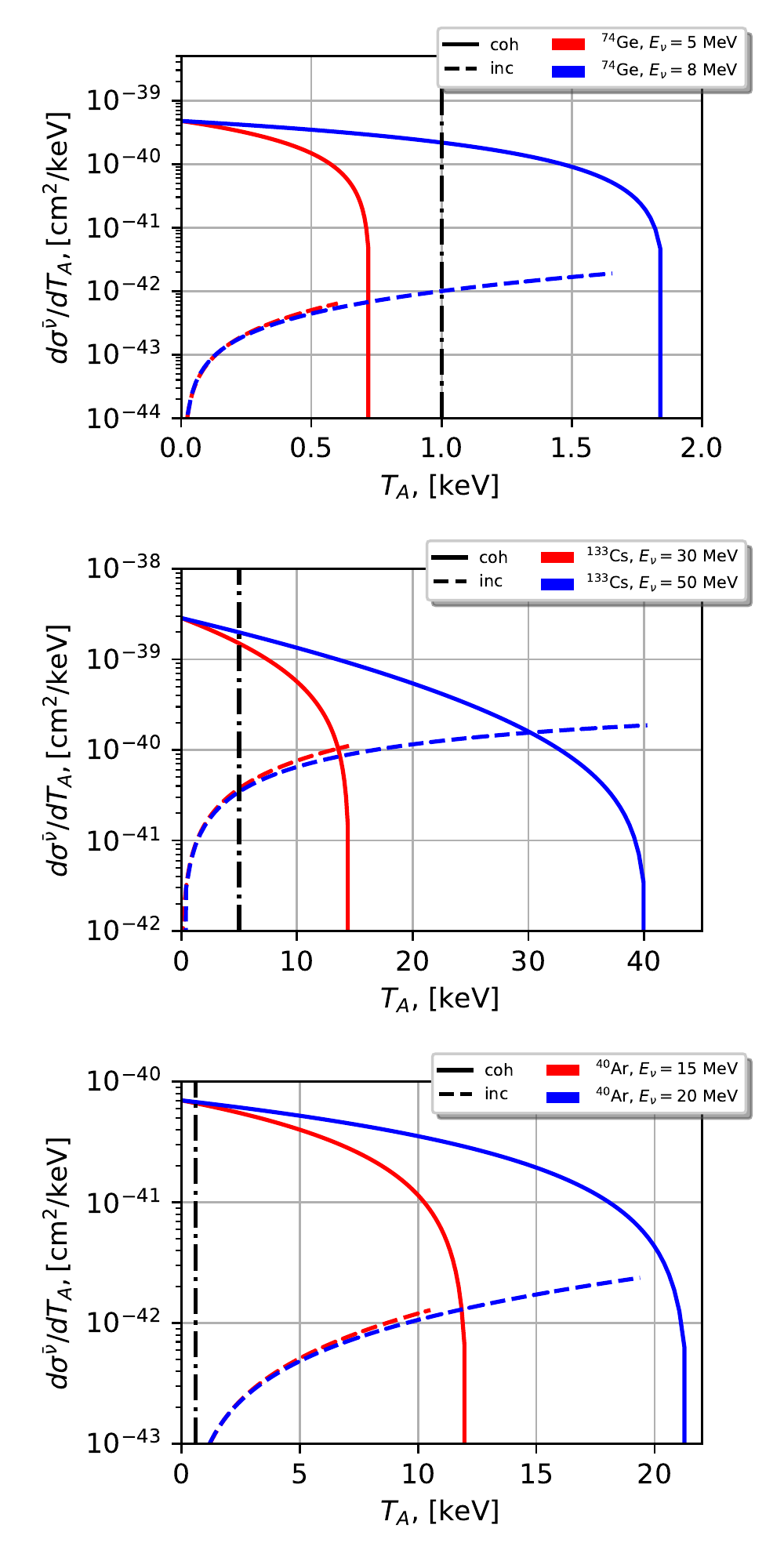}
\caption{Differential cross-sections $\displaystyle{\frac{d\sigma^{\nu/\bar\nu}}{dT_A}}$ for the coherent (solid lines) and incoherent (dashed lines) $\nu$-nucleus (left panel) and $\bar\nu$-nucleus (right panel) scattering for the ${}^{74}\text{Ge}$  (top), ${}^{133}\text{Cs}$ (middle), and ${}^{40}\text{Ar}$ (bottom) nuclei and different (anti)neutrino energies. Vertical lines correspond to the experimental energy thresholds.}
\label{fig:Xsection_differential-Nu+ANu}
\end{figure}

As in \cite{Bednyakov:2018mjd}, three experimental setups are considered.
The first is a germanium detector with the natural isotope ${}^{74}\text{Ge}$ only (for illustration), being exposed to the $\bar{\nu}_e$ flux from a nuclear reactor.
The energy threshold for the electrons of the Ge bolometers in 
the $\nu$GEN experiment at the Kalinin Nuclear Power Plant 
is 200 eV~\cite{nuGEN,Belov:2015ufh}, which roughly corresponds to 1 keV \cite{Barker:2012ek} of the ${}^{74}\text{Ge}$ recoil energy.
The differential cross-sections for $E_{\nu}=$ 5 MeV and 8 MeV and the total cross-section for $E_\nu\in (1,20)$ MeV were calculated.
As an estimate, $\Delta\varepsilon=900$ keV was used for the excitation energy of ${}^{74}\text{Ge}$.
The second setup is a CsI scintillator exposed to the neutrinos from the SNS~\cite{Akimov:2017ade}.
The differential and   total cross-sections are calculated for $E_\nu=30$ MeV and $50$ MeV and for $E_\nu\in (1,150)$ MeV, respectively.
It was assumed that $\Delta\varepsilon=100$ keV for the ${}^{133}\text{Cs}$ nucleus.
A 5-keV energy threshold was set to the ${}^{133}\text{Cs}$ recoil energy.
The third one is a liquid argon detector with an unprecedented low-energy threshold of $0.6$ keV for the ${}^{40}\text{Ar}$ nucleus achieved by the DarkSide Collaboration~\cite{Agnes:2018ves}.
The differential and   total cross-sections are calculated for $E_\nu=15(20)$ MeV and for $E_\nu\in (1,50)$ MeV, respectively.

In Fig.~\ref{fig:form-factors} the Helm form-factors for these nuclei as functions of $|\bm{q}|$ (and $T_{\rm Cs}$)  are depicted.
At $T_{\rm Cs}=$ 12--15 keV, where the maximum of the signal observed by the COHERENT experiment occurred, $|\bm{q}|= $ 50--60 MeV and $|F(\bm{q})|^2 =$ 0.6--0.5. 
It is seen that the coherent elastic scattering is suppressed, and a contribution from the incoherent transitions should be expected. 
\smallskip

\begin{figure}[p]
\includegraphics[width=0.50\linewidth]{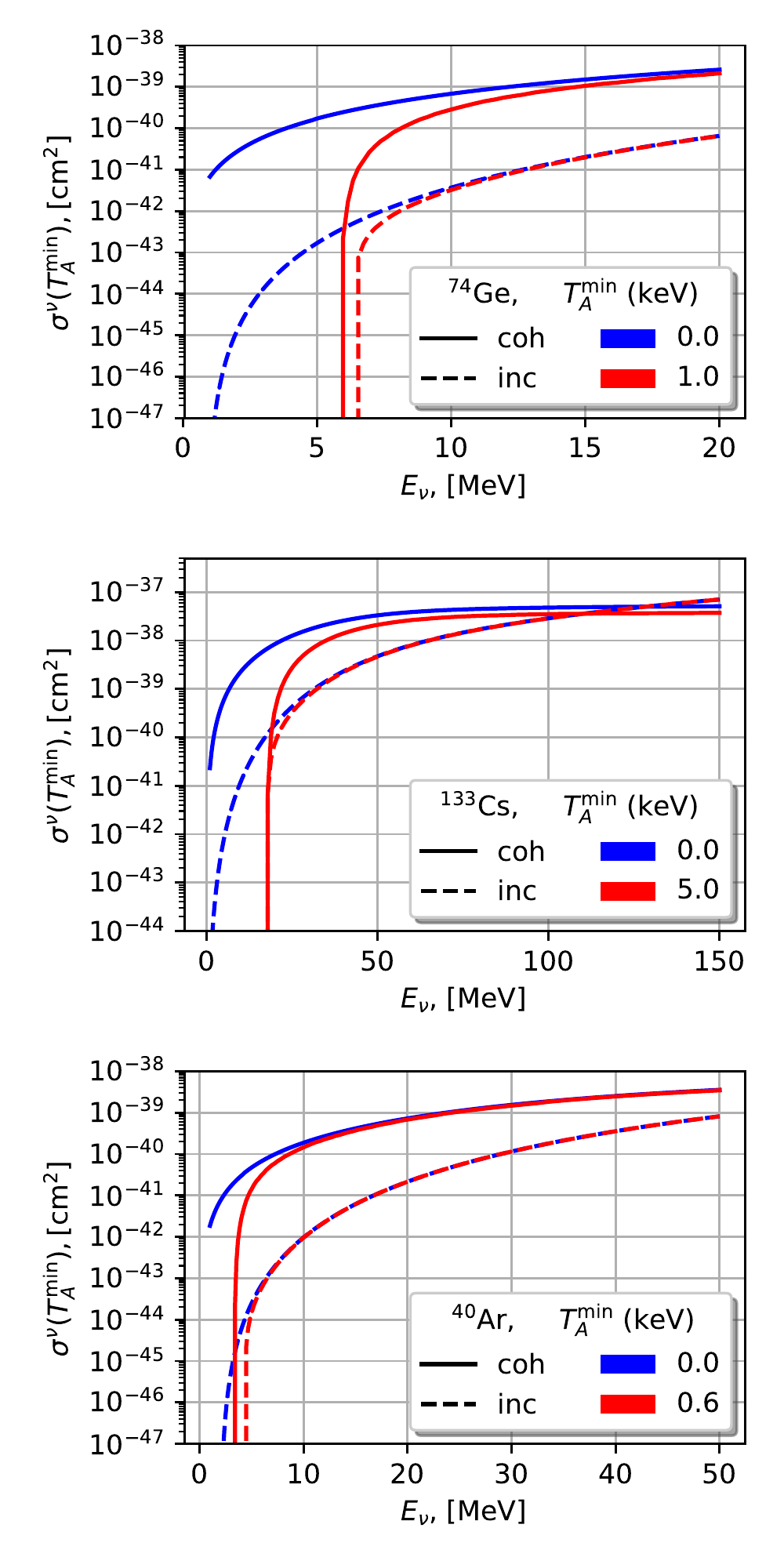}%
\includegraphics[width=0.50\linewidth]{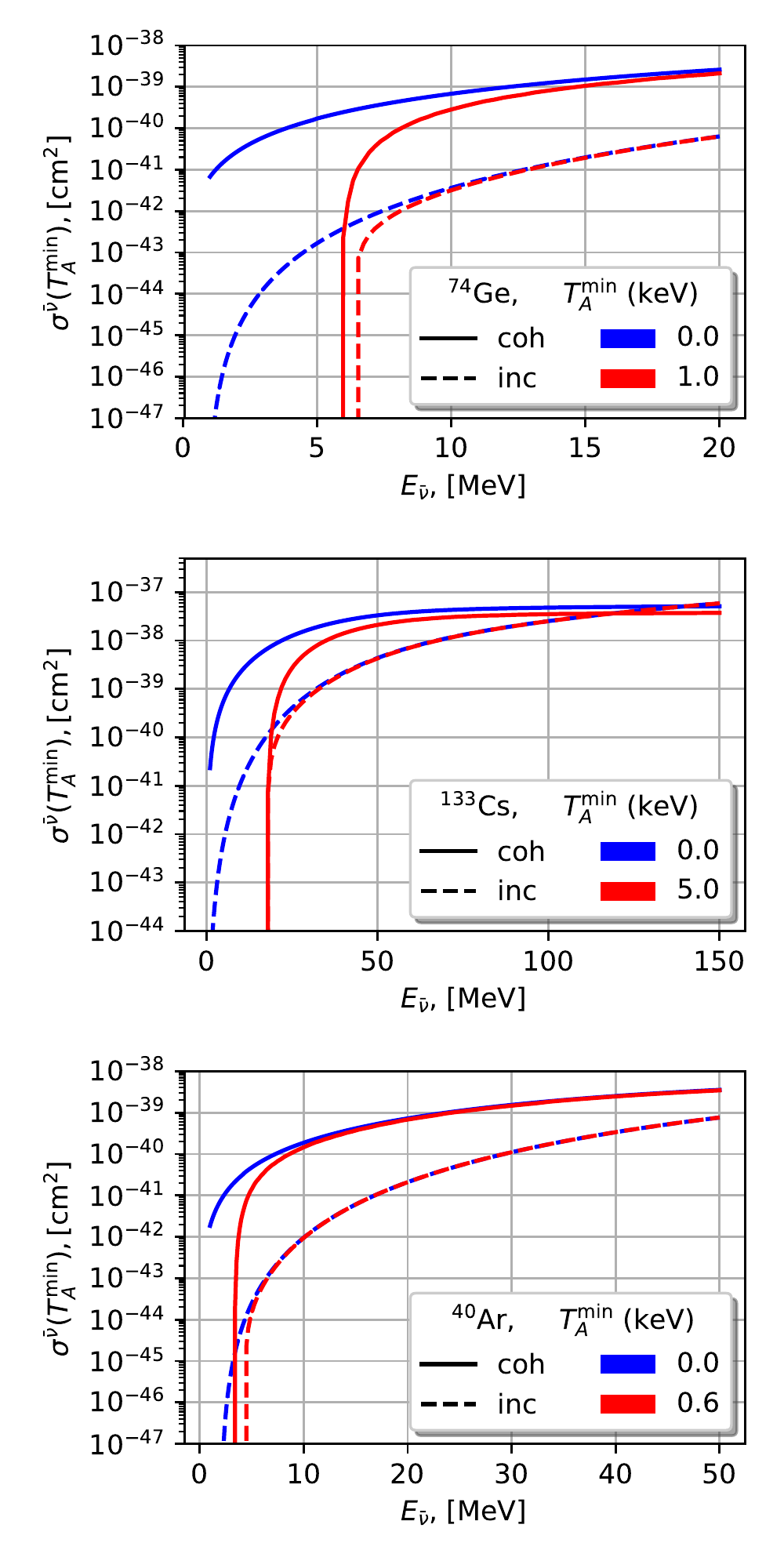}
\caption{Integral cross-sections $\sigma^{\nu/\bar\nu}$ for coherent (solid lines) and incoherent (dashed lines) $\nu$-nucleus (left) and $\bar\nu$-nucleus (right) scattering for  ${}^{74}\text{Ge}$  (top), ${}^{133}\text{Cs}$ (middle) and ${}^{40}\text{Ar}$ (bottom) nuclei and different (anti)neutrino energies.
The integrals are calculated for idealistic thresholdless ($T_A^\text{min}=0$, blue lines) setups and for the best achieved 
thresholds $T_A^\text{min}$ (red lines) achieved by three considered experimental setups.}
	\label{fig:Xsection_integral-Nu+ANu} 
\end{figure}

In~\cref{fig:Xsection_differential-Nu+ANu} (\cref{fig:Xsection_integral-Nu+ANu})
the differential (integral) coherent and incoherent (anti)neutrino-nucleus cross-sections are displayed for three experimental setups discussed above.

The following features can be seen in the figures.
The coherent and incoherent neutrino-nucleus and antineutrino-nucleus cross-sections, calculated with formulas (\ref{eq:sigma-inc1}) and (\ref{eq:sigma-coh1}), demonstrate, in accordance with discussion above, almost the same behavior
with a rather small difference only for the heavy non-spin-zero ${}^{133}\text{Cs}$ nucleus. 

As $T_A\to 0$, the coherent cross-section totally dominateos, since the incoherent contribution vanishes. 
As $T_A\to T_A^\text{max}$, the coherent cross-section vanishes due to the factor $1-T_A/T_A^\text{max}$, and the incoherent cross-section rises. 
Due to possible excitation of a nucleus the maximum kinetic energy of the nucleus 
in an incoherent process is systematically smaller than the one in the coherent case.
For small $E_{\nu/\bar\nu}$ the coherent cross-section dominates over the incoherent contribution for any $T_A$.
For larger $E_{\nu/\bar\nu}$ there is a value of $T_A$ above which the incoherent cross-section dominates over the coherent one, as can be seen in the middle panel of
~\cref{fig:Xsection_differential-Nu+ANu} for $E_{\nu/\bar\nu}=50$ MeV.
At low $E_{\nu/\bar\nu}$ the coherent integral cross-section (in \cref{fig:Xsection_integral-Nu+ANu}) is larger than the incoherent one by orders of magnitude because the factors $1-|F_{p/n}(\bm{q})|^2$ suppress the latter at small $\bm{q}$.
With increasing neutrino energy, their interrelation changes, and the integral incoherent cross-section becomes rather substantial above a certain $E_{\nu/\bar\nu}$.

\begin{figure}[!h]
\includegraphics[width=0.495\linewidth]{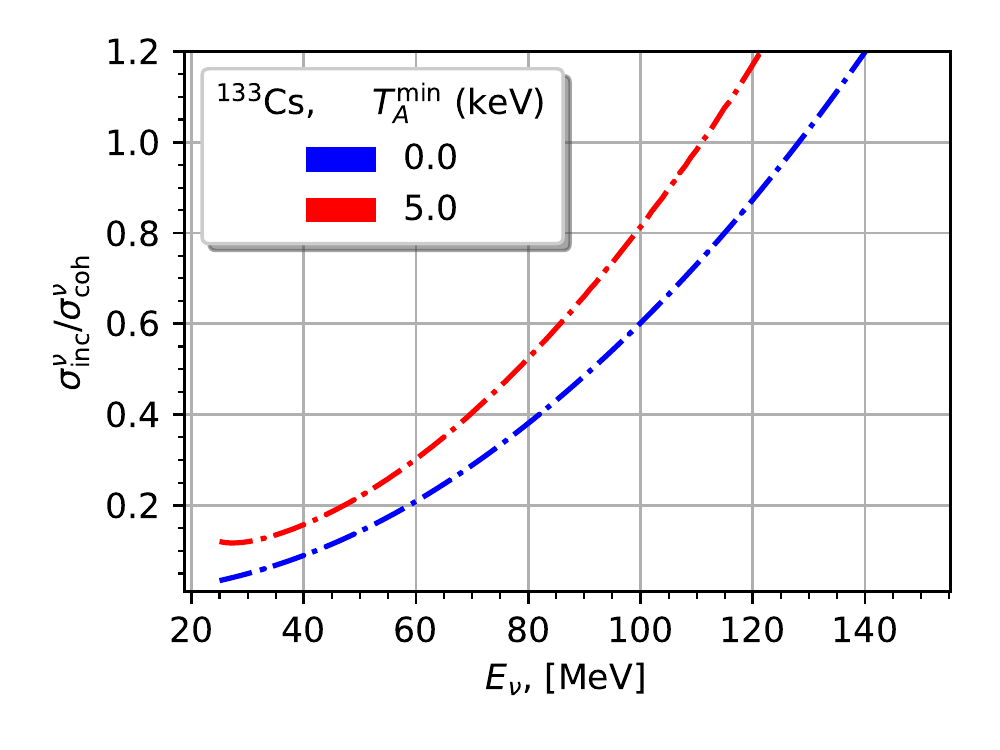} 
\includegraphics[width=0.495\linewidth]{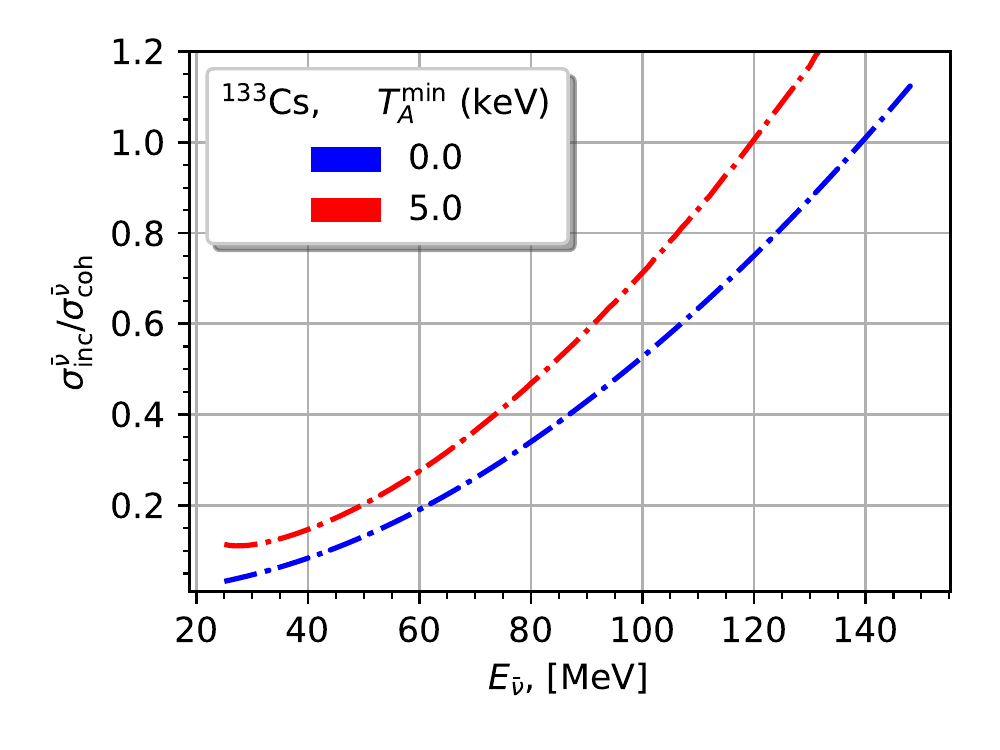}
\caption{Ratio of incoherent-to-coherent cross-sections, $\displaystyle{\sigma_\text{inc}/\sigma_\text{coh}}$, for  the neutrino (left) and antineutrino (right) scattering off a ${}^{133}\text{Cs}$ nucleus as a function of $E_{\nu/\bar\nu}$. 
The two curves correspond to the $T_A^\text{min}=0$ and 5 keV detection thresholds.}
\label{fig:Xsection_integral_ratios-Nu+ANu}
\end{figure}

Figure~\ref{fig:Xsection_integral_ratios-Nu+ANu} illustrates this statement. 
The ratio $\sigma_\text{incoh}/\sigma_\text{coh}$ of the integrals given by~\cref{eq:integral_xs} 
 is displayed for the ${}^{133}\text{Cs}$ nucleus.
For the neutrino-nucleus scattering (left panel) and $E_\nu\simeq 30 \ (60)$ MeV  
this ratio is about 7 (20)\% for $T_A^\text{min}=0$, and reaches about 15 (30)\% for $T_A^\text{min}=5$ keV.
In the latter case, the incoherent contribution becomes equal to the coherent one at $E_{\nu}\simeq$ 110 MeV.  
The increasing importance of the incoherent interaction is evident for increasing neutrino energy.

After the interaction with incoming (anti)neutrino the nucleus may remain in the same quantum state, or the internal state of the nucleus could be changed. 
Experimentally, the scattered nucleus, being in the same or excited state, is practically indistinguishable if one measures only the kinetic energy of the nucleus.
Nevertheless, inelastic interaction (for example, nuclear excitation) must be accompanied by 
some emission of $\gamma$ quanta corresponding to the difference of the energy levels of the nucleus~\cite{Donnelly:1975ze}.
For example, the time scale of this emission is in the range of picoseconds to nanoseconds for the ${}^{133}\text{Cs}$ nucleus.
The energies of the $\gamma$s are of the order of a hundred keV for ${}^{133}\text{Cs}$, and 
these $\gamma$s should produce a very detectable signal in the scintillator correlated in time with the beam pulses for an accelerator-based experiment.
The rate of these $\gamma$s is determined by the ratio $N_\text{inc}/N_\text{coh}$, where
$$N_\text{inc/coh} = \int dE_\nu \Phi(E_\nu)\int_{dT_A^\text{min}}^{dT_A^\text{max}} dT_A\frac{d\sigma_\text{inc/coh}}{dT_A} \varepsilon(T_A),
$$ and $\varepsilon(T_A)$ is the detection efficiency.
Figure \ref{fig:Xsection_integral_ratios-Nu+ANu} suggests that the number of the $\gamma$-events due to incoherent interactions could be rather detectable.

One can conclude that  the COHERENT experiment (with ${}^{133}\text{Cs}$) has seen 
a very substantial part of Coherent Elastic Neutrino Nucleus Scattering (CE$\nu$NS), but with 15--20\% uncertainty, 
due to the high neutrino energy and the high energy threshold (5 keV).
The inelastic (or incoherent) admixture at a level of 15-20\% is inevitable in the measured data of the experiment, 
if excitation $\gamma$s escape detection.  
An accurate analysis of the COHERENT-like data (see for example
\cite{Cadeddu:2018izq,Boehm:2018sux,Brdar:2018qqj,Cadeddu:2018dux,Millar:2018hkv,Altmannshofer:2018xyo,%
Blanco:2019vyp,AristizabalSierra:2019zmy,Huang:2019ene,Miranda:2019skf,Papoulias:2019lfi}) 
should take the incoherent contribution into account. 

There are two ways for an accurate study of the CE$\nu$NS.
One is to separate the coherent signal from the incoherent one following the above-mentioned procedure from  
\cite{Bednyakov:2018mjd}. 
The incoherent processes, being a relatively small ''background'' to the coherent interactions, provide an important clue if $\gamma$ rays emitted by the excited nucleus are detected.
For a neutrino pulsed-beam experiment the $\gamma$s should be correlated in time with the beam pulse, and the higher energy of the $\gamma$s allows their detection at a rate governed by the ratio  $N_\text{inc}/N_\text{coh}$.
Simultaneous detection of both signals due to nuclear recoil and the deexcitation $\gamma$s provides a sensitive tool for investigation of the CE$\nu$NS, and studies of the nuclear structure and possible signs of new physics.
The other way to study the CE$\nu$NS is to use an extremely low-energy threshold detector and collect data at recoil energies, where the incoherent (inelastic) scattering is suppressed very significantly.
Nowadays, this is an objective for the $\nu$GeN experiment 
\cite{nuGEN,Belov:2015ufh} and, perhaps in the near future, for the DarkSide experiment \cite{Agnes:2018ves} with their very-low-energy thresholds. 

Some comments are in order after publications of \cite{Bednyakov:2018mjd}.
Analytical treatment of the neutrino-nucleon scattering within a nucleus was possible under a number of approximations and assumptions.
In particular, non inclusion of the nucleon spin-flip transitions into the elastic process was an assumption in \cite{Bednyakov:2018mjd}. 
This spin-flip transition could lead to any of elastic and inelastic processes.
The corresponding probabilities are determined by the nucleus wave function, unlikely to be calculable from first principles.
Furthermore, any spin-flip of a target nucleon in a spinless nucleus would necessarily change the total spin of the nucleus and thus excite it.
Therefore, it is likely that nuclei with $A\gg \Delta A$ would behave similarly, i.e. the nucleon spin-flip transitions will change the energy state of the nucleus, etc.

Another comment concerns the point that according to tradition,  
the cross-sections in Eqs. (\ref{eq:main-cross-section}), 
(\ref{eq:sigma-inc1}) and (\ref{eq:sigma-coh1}) have labels 
"coherent"\/ and "incoherent"\/. 
Nevertheless, for (anti)neutrino energies of tens of MeV both  
terms in Eq. (\ref{eq:sigma-inc1}) and Eq. (\ref{eq:sigma-coh1})
are coherent in the sense that all nucleons are involved in the scattering process at a level of amplitudes 
(for details see \cite{Bednyakov:2018mjd}).
Therefore, strictly speaking in this kinematics region one should use 
labels "elastic"\/ and "inelastic"\/ instead of "coherent"\/ and "incoherent"\/, 
respectively.
For much higher energies (tens of GeV and beyond) one can, in principle, generalize the excitation of the target nucleus $\nu A\to \nu A^*$ considered here to a scattering process when the nucleus is fully disintegrated  
$\nu A\to \nu X$ (deep inelastic case). 
The full disintegration obviously means that the scattered nucleus completely 
lost its integrity, or equivalently $F(\bm{q})\equiv 0$. 
The partonic picture could be foreseen  
from Eq. (\ref{eq:sigma-inc1}) with its famous $A$-dependence.

In conclusion, a unified description of the elastic (coherent) and inelastic (incoherent)  neutrino and antineutrino scattering off the nucleus is presented. This description can be used for comprehensive data analysis.     

The authors are grateful to Yu.~Efremenko, S.~Haselschwardt, A.~Konovalov, V.~Rubakov, and E.~Yakushev,   
for important comments and discussions.

\bibliographystyle{apsrev4-1}
\bibliography{COHERENT}

\end{document}